\documentclass{birkmult}
 \newtheorem{thm}{Theorem}[section]

 \theoremstyle{definition}
 
 \theoremstyle{remark}

 \newcommand{\R}{\mathbb{R}}
 \newcommand{\Hy}{\mathbb{H}}
 \newcommand{\T}{\textstyle}
 \newcommand{\ul}[1]{\underline{#1}}
 \numberwithin{equation}{section}

\begin{document}
%
%
%
%
%
%
%
%
%
\title[IR Problem in AdS/CFT]
 {A Comment on the Infra-Red Problem\\  in the AdS/CFT Correspondence}
\author[Gottschalk]{Hanno Gottschalk}

\address{%
IAM, Wegelerstr. 6\\
D-53115 Bonn\\
Germany}

\email{gottscha@wiener.iam.uni-bonn.de}

\author[Thaler]{Horst Thaler}
\address{Dipartimento di Matematica e Informatica\br
 Universit\`{a} di Camerino\br Italy} \email{horst.thaler@unicam.it}


\date{October 15, 2007}

\begin{abstract}
In this note we report on some recent progress in proving the
AdS/CFT correspondence for quantum fields using rigorously defined
Euclidean path integrals. We also comment on the infra-red  problem
in the AdS/CFT correspondence and argue that it is different from
the usual IR problem in constructive quantum field theory. To
illustrate this, a triviality proof based on hypercontractivity
estimates is given for the case of an ultra-violet regularized
potential of type $:\phi^4:$. We also give a brief discussion on
possible renormalization strategies and the specific problems that
arise in this context.
\end{abstract}

\maketitle
\section{Introduction}
Often, the AdS/CFT correspondence between string theory or some
other theory including quantized gravity on bulk AdS and
super-symmetric Yang-Mills theory on its conformal boundary
\cite{Ma,Wi} is formulated in terms of Euclidean path integrals. In
the absence of mathematically rigorous approaches to path integrals
of string type (see however \cite{AJPS}) or even gravity, it seems
to be reasonable to use the well-established theory of constructive
quantum field theory (QFT) \cite{GJ} as a testing lab for some
aspects of the more complex original AdS/CFT conjecture. That such
simplified versions of the AdS/CFT correspondence are in fact
possible was already noted by Witten \cite{Wi} (see also \cite{GKP})
and further elaborated by \cite{DR}. In \cite{GT}  we give a
mathematically rigorous version of the latter work (in \cite{Ha} one
finds some related ideas), leaving however the infra-red (IR)
problem open. In this note we come back to the IR problem and we
show how the difference between the IR problem in the AdS/CFT
correspondence as compared with the usual IR problem in constructive
QFT leads to somewhat unexpected results.

The authors would like to underline that, in contrast to \cite{GT},
the present article is rather focused on ideas and thus leaves space
for the interpretation of the validity of the results. We will
comment on that in several places.

The article is organized as follows: In the following section we
introduce the mathematical framework of AdS/CFT correspondence and
define rigorous probabilistic path integrals on AdS. In Section
\ref{3sect} we recall the main results from \cite{DR,GT}, i.e. that
the generating functional that is obtained from imposing certain
boundary conditions at the conformal boundary (which is the way
generating functionals are defined in string theory) can in fact be
written as a usual generating functional of some other field theory.
From the latter form it is then easy to extract structural
properties, e.g. reflection positivity of the functional, in the
usual way. Somewhat unexpectedly, it is not clear whether a
functional integral can be associated to the boundary theories.
These statements hold for all sorts of interactions with a
IR-cut-off. In Section \ref{4sect} the IR-problem in this version of
the AdS/CFT correspondence is discussed on a heuristic level. We
also sketch the proof of triviality of the generating functional of
the conformally invariant theory on the conformal boundary of AdS
for the case of an UV-regularized $:\phi^4:$ interaction.  We
briefly survey strategies that might be candidates to overcome the
triviality obstacle at a non-rigorous level and we comment on
specific problems with such strategies. The final section gives some
preliminary conclusions and an outlook on open research problems in
understanding further the mathematical basis of AdS/CFT.

\section{Functional integrals on AdS}
\label{2sect}

 Let us consider the $d+2$ dimensional ambient space
$\R^{d,2}=\R^{d+2}$ with inner product of signature
$(-,+,\ldots,+,-)$, i.e.
$\zeta^2=-\zeta_1^2+\zeta_2^2+\cdots+\zeta_{d+1}^2-\zeta_{d+2}^2$
where $\zeta\in\R^{d,2}$. Then the submanifold defined by
$\{\zeta\in\R^{d,2}: \zeta^2=-1\}$ is a $d+1$ dimensional Lorentz
manifold with metric induced by the ambient metric. It is called the
$d+1$ dimensional Anti de Sitter (AdS) space. Formal Wick rotation
$\zeta_1\rightarrow i\zeta_1$ converts the ambient space into the
space $\R^{d+1,1}$ with signature $(+,\ldots,+,-)$. Under Wick
rotation, the AdS space is converted to the Hyperbolic space
$\Hy^{d+1}:\{\zeta\in\R^{d+1}:\zeta^2=-1, \zeta^d>0\}$, which is a
Riemannian submanifold of the ambient $d+2$ dimensional Minkowski
space. We call $\Hy^{d+1}$ the Euclidean AdS space.

It has been established with full mathematical rigor that Euclidean
random fields that fulfill the axioms of invariance, ergodicity and
reflection positivity give rise, via an Osterwalder--Schrader
reconstruction theorem, to local quantum field theories on the
universal covering of the relativistic AdS, cf. \cite{BEM,JR}
justifying the above sketched formal Wick rotation. Hence a
constructive approach with reflection positive Euclidean functional
integrals is viable.

It is convenient to work in the so called half-space model of
Euclidean AdS (henceforth the word Euclidean will be dropped). This
coordinate system is obtained via the change of variables
$\zeta_i=x_i/z$, $i=1,\ldots,d$, $\zeta_{d+1}=-(z^2+x^2-1)/2z$,
$\zeta_{d+2}=(z^2+x^2+1)/2z$ which maps
$\R_+^{d+1}=\{(z,x)\in\R^{d+1}: z>0\}$ to $\Hy^{d+1}$. We will use
the notation $\ul{x}$ for $(z,x_1,\ldots,x_d)\in \R^{d+1}_+$. The
metric on $\R^{d+1}_+$ is given by
$g=(dz^2+dx_1^2+\cdots+dx_d^2)/z^2$ which implies that the
canonical volume form is $d_g\ul{x}=z^{-d-1}dz\wedge d
x_1\wedge\cdots\wedge dx_d$. The conformal boundary of $\Hy^{d+1}$
then is the $d$-dimensional Euclidean space $\R^d$ with metric
$ds^2=dx_1^2+\cdots+dx_d^2$ which is obtained via the limit $z\to 0$
and a conformal transformation of the AdS metric. Of course, the
upshot of the AdS/CFT correspondence is that the action of the
Lorentz group on the AdS space $\Hy^{d+1}$ gives rise to an action of
the conformal group transformations on the conformal boundary. One
thus expects an AdS symmetric QFT (or string/quantum gravity\ldots
theory) on the bulk $\Hy^{d+1}$ to give, if properly restricted to
the conformal boundary, a conformally invariant theory on $\R^{d}$.

We will now make this precise. On the hyperbolic space $\Hy^{d+1}$
one has  two invariant Green's functions (``bulk-to-bulk
propagators") for the operator $-\Delta_g+m^2$, with $\Delta_g$ the
Laplacian and $m^2$ a real number suitably bounded from below, that
differ by scaling properties towards the conformal boundary
\begin{equation}
\label{2.1eqa}
G_\pm(z,x;z',x')=\gamma_{\pm}(2u)^{-\Delta_{\pm}}F(\Delta_{\pm},\Delta_{\pm}+{\T
\frac{1-d}{2}};2\Delta_{\pm} +1-d;-2u^{-1})
\end{equation}
Here $F$ is the hypergeometric function,
$u=\frac{(z-z')^2+(x-x')^2}{2zz'}$,
$\Delta_{\pm}=\frac{d}{2}\pm\frac{1}{2}
\sqrt{d^2+4m^2}$ \newline $=:\frac{d}{2}\pm\nu, \nu> 0$ and $\gamma_\pm
=\frac{\Gamma(\Delta_\pm)}{2\pi^{d/2}\Gamma(\Delta_\pm+1-\frac{d}{2})}$
\cite{DR,GT}.  Taking pointwhise scaling limits for $z\to 0$ in one
or two of the arguments, the bulk-to-boundary and
boundary-to-boundary propagators are obtained
\begin{equation}
\label{2.2eqa} H_\pm(z,x;x')=\lim_{z'\rightarrow
0}z'^{-\Delta_\pm}G_\pm(z,x;z',x')
=\gamma_\pm\left(\frac{z}{z^2+(x-x')^2}\right)^{\Delta_\pm}
\end{equation}
and
\begin{equation}
\label{2.3eqa} \alpha_\pm(x,x')=\lim_{z \rightarrow 0
}z^{-\Delta_\pm}H_\pm(z,x;x') =\gamma_\pm(x-x')^{-2\Delta_\pm}.
\end{equation}
If (\ref{2.2eqa}) or (\ref{2.3eqa}) do not define locally integrable
functions, the expressions on the right hand side are defined via
analytic continuation in the weights $\Delta_\pm$. An important
relation between $G_+$, $G_-$, $H_+$ and $\alpha_-$ is the
covariance splitting formula for $G_-$ given by
\begin{equation}
\label{2.4eqa} G_-(\ul{x},\ul{x}')= G_+(\ul{x},\ul{x}') +
\int_{\mathbb{R}^d}\int_{\mathbb{R}^d}
 H_+(\ul{x},y)c^2\alpha_-(y,y')H_+(\ul{x}',y')dy dy',
 \end{equation}
with $c=2\nu$.

 We now pass on to the description of mathematically
well-defined functional integrals. Let ${\mathcal D}={\mathcal
D}(\Hy^{d+1},\mathbb{R})$ be the infinitely differentiable,
compactly supported functions on $\Hy^{d+1}$ endowed with the
topology of compact convergence. The propagator $G_+$ is the
resolvent function to the Laplacian $\Delta_g$ with Dirichlet
boundary conditions at conformal infinity, from which it follows
that $G_+$ is stochastically positive, $\langle
f,f\rangle_{-1}=G_+(\bar
f,f)=\int_{\Hy^{d+1}\times\Hy^{d+1}}G_+(\ul{x},\ul{x}')\bar
f(\ul{x})f(\ul{x}') \, d_g\ul{x} d_g\ul{x}'\geq 0$ $\forall
f\in{\mathcal D}$,
and reflection positive as
long as $m^2>-\frac{d^2}{4}$. The latter value is determined by the lower
bound of the spectrum of $\Delta_g$ on $\Hy^{d+1}$. In explicit, if
$\theta: (z,x_1,x_2,\ldots,x_d)\to(z,-x_1,x_2,\ldots,x_d)$ is the
reflection in $x_1$-direction, then for any $f\in{\mathcal
D}_+=\{h\in {\mathcal D}:h(\ul{x})=0$ if $x_1\leq 0\}$ we have
$$\int_{\Hy^{d+1}\times\Hy^{d+1}}G_+(\ul{x},\ul{x}') \bar
f_\theta(\ul{x})f(\ul{x}') \, d_g\ul{x} d_g\ul{x}'\geq 0,$$ cf. \cite{GJ}.
Here, $f_u(\ul{x})=f(u^{-1}\ul{x})$ for $u\in{\rm Iso}(\Hy^{d+1})$.

Consequently, via application of Minlos theorem, there exists a
unique probability measure $\mu_{G_+}$ on the measurable space
$({\mathcal D}',{\mathcal B})$, where ${\mathcal D}'$ is the
topological dual space of ${\mathcal D}$ and ${\mathcal B}$ the
associated Borel sigma algebra, such that $\int_{{\mathcal
D}'}e^{\langle \phi,f\rangle}d\mu_{G_+}(\phi)=e^{\frac{1}{2}\langle
f,f\rangle_{-1}}$. By setting $\varphi(f)(\phi)=\phi(f)$ we define
the canonical random field associated with $\mu_{G_+}$, i.e. a
random variable valued distribution. In the following we omit the
distinction between $\varphi$ and $\phi$ and write $\phi$ for both.

Let ${\mathcal B}_\Lambda$, $\Lambda\subseteq\Hy^{d+1}$ be the
smallest sigma algebra generated by the functions ${\mathcal
D}'\ni\phi\to\langle\phi,f\rangle$, ${\rm supp}f\in\Lambda$ and
$M(\Lambda)$ be the functions that are ${\mathcal
B}_\Lambda$-measurable. We use the special abbreviations ${\mathcal
B}_+={\mathcal B}_{\{\ul{x}\in\Hy^{d+1}:x_1>0\}}$ and
$M_+=M({\mathcal B}_+)$. Then $\mu_{G_+}$ is reflection positive,
i.e.
\begin{equation}
\label{2.5eqa} \int_{{\mathcal D}'}\Theta\bar F(\phi)F(\phi)\,
d\mu_{G_+}(\phi)\geq 0,~~\forall F\in M_+.
\end{equation}
The reflection $\Theta F(\phi)$ is defined as $F(\phi_\theta)$ with
$\langle\phi_u,f\rangle=\langle\phi, f_{u^{-1}}\rangle$ $\forall
\phi\in{\mathcal D}'$, $f\in{\mathcal D}$ and $u\in{\rm
Iso}(\Hy^{d+1})$. $\langle\,.\,,\,.\,\rangle$ is the duality between
${\mathcal D}'$ and ${\mathcal D}$ induced by the
$L^2(\Hy^{d+1},d_g\ul{x})$ inner product.

Let $\{V_\Lambda\}:{\mathcal D}'\to\R$ be a set of interaction
potentials indexed by the net of bounded, measurable subsets
$\Lambda$ in $\Hy^{d+1}$. In particular these sets have finite
volume $|\Lambda|=\int_\Lambda d_g\ul{x}$. We require that the
following conditions hold:
\begin{itemize}
\item[(i)] Integrability: $e^{-V_\Lambda}\in L^1({\mathcal
D}',d\mu_{G_+})$ $\forall \Lambda$;
\item[(ii)] Locality: $V_\Lambda\in M({\mathcal B}_\Lambda)$;
\item[(iii)] Invariance: $V_\Lambda(\phi_u)=V_{u^{-1}\Lambda}(\phi)$ $\mu_{G_+}$--a.s..
\item[(iv)] Additivity:
$V_\Lambda+V_{\Lambda'}=V_{\Lambda\cup\Lambda'}$ for
$\Lambda\cap\Lambda'=\emptyset$.
\item[(v)] Non-degeneracy: $V_\Lambda=0$ $\mu_{G_+}$--a.s. if $|\Lambda|=0$.
\end{itemize}
Then, using (i), we obtain a family of interacting measures on
$({\mathcal D}',{\mathcal B})$, indexed by the net $\{\Lambda\}$, by
setting $d\mu_{G_+,\Lambda}=e^{-V_\Lambda}d\mu_{G_+}/Z_\Lambda$ with
$Z_\Lambda=\int_{{\mathcal D}'}e^{-V_\Lambda}\,d\mu_{G_+}$.
Furthermore, using (ii)--(v) we get whenever $\theta
\Lambda=\Lambda$
\begin{equation}
\label{2.6eqa} \int_{{\mathcal D}'}\Theta\bar F F\,
d\mu_{G_+,\Lambda}=\frac{1}{Z_\Lambda}\int_{{\mathcal
D}'}\Theta\left(\bar
Fe^{-V_{\Lambda_+}}\right)\left(Fe^{-V_{\Lambda_+}}\right)
d\mu_{G_+}\geq 0,~~\forall F\in M_+,
\end{equation}
where $\Lambda_+=\Lambda\cap\{x\in\Hy^{d+1}:x_1>0\}$. Hence
reflection positivity is preserved under the perturbation.
Furthermore, from the invariance of $\mu_{G_+}$ under ${\rm
Iso}(\Hy^{d+1})$ we get that
$u_*\mu_{G_+,\Lambda}=\mu_{G_+,u\Lambda}$. Here $u\in {\rm
Iso}(\Hy^{d+1})$ induces an action on ${\mathcal D}'$ via
$\phi\to\phi_u$ and $u_*$ is the pushforward under this action.
Consequently, if the limit (in distribution)
$\mu_{G_+,\Hy^{d+1}}=\lim_{\Lambda\nearrow\Hy^{d+1}}\mu_{G_+,\Lambda}$
exists and is unique, the limiting measure is invariant under ${\rm
Iso}(\Hy^{d+1})$ and reflection positive. Invariance follows from
the equivalence of the nets $\{\Lambda\}$ and $\{u\Lambda\}$ and the
postulated uniqueness of the limit over the net $\{\Lambda\}$.

Let us next consider functional integrals associated with the
Green's function $G_-$. In the case when $2\nu<d$ ($\Leftrightarrow
m^2<0$)  we get that $\alpha_-$ is stochastically positive since
$\alpha_-(\bar f,f)=\int_{\R^d\times\R^d}\alpha_-(x,x')\bar
f(x)f(x') \, dxdx'=C_{-\nu}\int_{\R^d}|k|^{-2\nu}|\hat
f(k)|^2\,dk\geq0$. $\hat f$ denotes the fourier transform of $f$
wrt $x$, $\hat f(k)=(2\pi)^{-d/2}\int_{\R^d}e^{ik\cdot x}f(x)\,
dx$. Furthermore, $\alpha_-$ is reflection positive in
$x_1$-direction (in the usual sense, cf. \cite{GJ}) if and only if
$-\nu>-1$, which is also known as the unitarity bound. It is clear
from the decomposition (\ref{2.4eqa}) that $G_-$ is stochastically
positive if $G_+$ and $\alpha_-$ are both stochastically positive. The
reflection positivity of $G_-$ does not follow from the reflection
positivity of $G_+$ and $\alpha_-$ due to the non-local effect of
$H_+$. We will however not need it here. We thus conclude that for
$\sup {\rm spec}({\Delta_g})<m^2<0$ a unique probability measure
$\mu_{G_-}$ on $({\mathcal D}',{\mathcal B})$ with Laplace transform
$\int_{{\mathcal
D}'}e^{\langle\phi,f\rangle}d\mu_{G_-}(\phi)=e^{\frac{1}{2}\langle
f,f\rangle_{-1,-}}$ exists. Here $\langle
f,f\rangle_{-1,-}=G_-(f,f)$. The perturbation of $\mu_{G_-}$ with an
interaction can now be discussed in analogy with the above case --
where however the reflection positivity for the perturbed measure
remains open, as reflection positivity of the free measure does not
necessarily hold.

\section{Two Generating Functionals}
\label{3sect} On the string theory side of the AdS/CFT
correspondence, generating functionals for the boundary theory are
calculated fixing boundary conditions at the conformal boundary (so
called Dirichlet boundary conditions). Little is known about the
mathematical properties of such kinds of generating functionals.
E.g. their stochastic and reflection positivity is far from obvious,
leaving the linkage to path integrals and relativistic physics open.
It was noticed by D\"utsch and Rehren \cite{DR} that such kinds of
generating functionals can however be re-written in terms of
ordinary generating functionals, from which the structural
properties can be read of in the usual way. These ideas in \cite{GT}
have been made fully rigorous in the context of constructive QFT. We
will now briefly review these results.

The generating functional $Z(f)/Z(0)$, $f\in{\mathcal S}(\R^d,\R)$,
the space of Schwartz functions, in the AdS/CFT correspondence from
a string theoretic point of view can be described as follows: Let
$\phi$ be some scalar quantum field that is included in the theory
(e.g. the dilaton field) and let $V_\Lambda$ be the (IR and
eventually UV-regularized) effective potential for that field
obtained via integrating out the remaining degrees of freedom
(leaving open the question how such an ``integral" can be defined).
To simplify the model and for the sake of concreteness we will
sometimes assume that $V_\Lambda$ is of polynomial type. Formally,
\begin{equation}
\label{3.1eqa} Z(f)=\int_{\phi_0=\phi|_{\partial \Hy^{d+1}}=f }
e^{-S_0(\phi)-V_\Lambda(\phi)}\, d\phi=\int \delta(\phi_0-f)
e^{-S_0(\phi)-V_\Lambda(\phi)}\, d\phi
\end{equation}
where $S_0=|\nabla\phi|^2+m^2\phi^2$, $\phi_0=\phi|_{\partial
\Hy^{d+1}}$ are suitably rescaled boundary values of the field
$\phi$ and $d\phi$ is the heuristic flat measure on the space of all
field configurations. The first step in making this formal
expression rigorous is to replace $e^{-S_0(\phi)}\, d\phi$ with a
well-defined probabilistic path integral. It turns out that
$d\mu_{G_-}(\phi)$ is the right candidate and hence for the moment
restriction to $m^2<0$ is necessary.

In a second step we have to make sense out of the boundary condition
$\phi_0=f$ or the functional delta distribution on the boundary
values of the field, respectively. Using the covariance splitting
formula (\ref{2.4eqa}) we obtain the splitting
$\phi_-(\ul{x})=\phi_+(\ul{x})+\int_{\R^d}H_+(\ul{x},x')\phi_{\alpha_-}(x')\,
dx'$, where $\phi_\pm$ are the canonical random fields associated
with $G_\pm$ and $\phi_{\alpha_-}$ is the canonical random field
associated to the functional measure $\mu_{\alpha_-}$, i.e. the
Gaussian measure with generating functional
$e^{\frac{1}{2}\alpha_-(f,f)}$ living on the conformal boundary of
$\Hy^{d+1}$.

The following step is to construct a finite dimensional
approximation $\psi_{\alpha_-}$ of the boundary field $\phi_{\alpha_-}$ by
projecting it via a basis expansion to $\R^n$. Thereafter, one can
implement the delta distribution as a delta distribution on $\R^n$.
Finally one can remove the finite dimensional approximation via a
limit $n\to\infty$. It turns out that this limit exists and is
unique up to a diverging multiplicative constant. This constant
however drops out in the quotient $Z(f)/Z(0)$. With the projection
to the first $n$ terms of the basis expansion denoted by $p_n$ and
$\eta$ a linear mapping from this space to $\R^n$ we get
$$
C_{A_{-}}\int_{\mathbb{R}^n} \int_{{\mathcal D}'}
\delta(\psi_{\alpha_-}-\eta p_n f)e^{-V_\Lambda(\phi_+ + c
H_+(\eta^{-1}\psi_{\alpha_-}))}d\mu_{G_+}(\phi_+)
e^{-\frac{1}{2}(\psi_{\alpha_-},A_-\psi_{\alpha_-})}d\psi_{\alpha_-}
$$
\begin{equation}\label{3.2eqa}
=C_{A_{-}}e^{-\frac{1}{2}(f,(p_n\alpha_- p_n)^{-1}f)}\int_{{\mathcal
D}'} e^{-V_\Lambda(\phi_+ + c H_+(p_n
f))}d\mu_{G_+}(\phi_+)=:Z_{n}(f),
\end{equation}
where $A_-:=(\eta p_n \alpha_{-} p_n \eta^{-1})^{-1}$ and
$C_{A_-}=\frac{|{\rm det}A_-|^{\frac{1}{2}}}{(2\pi)^{\frac{d}{2}}}$.
One can then show that
\begin{equation}
\label{3.3eqa} Z(f)/Z(0):=\lim_{n\to\infty}
Z_n(f)/Z_n(0)=e^{-\frac{1}{2}(f,\alpha_{-}^{-1}f)}\frac{\int_{{\mathcal
D'}}e^{-V_\Lambda(\phi_++c H_+f)}d\mu_{G_+}(\phi_+)}{\int_{{\mathcal
D'}}e^{-V_\Lambda(\phi_+)}d\mu_{G_+}(\phi_+)}
\end{equation}
converges under rather weak continuity requirements on $V_\Lambda$
that are fulfilled e.g. for UV-regularized potentials in arbitrary
dimension and for $P(\phi)_2$ potentials without UV cut-offs in
$d+1=2$. Obviously, the limit does not depend on the details of the
finite dimensional approximation. For the details we refer to
\cite{GT}. We now realize that the right hand side of (\ref{3.3eqa})
also makes sense for $m^2\geq 0$ and we adopt (\ref{3.3eqa}) as a
definition of (\ref{3.1eqa}).

At this point one would like to associate a boundary field theory to
the generating functional ${\mathcal C}(f)=Z(f)/Z(0)$. In order to
obtain a functional integral associated to ${\mathcal C}:{\mathcal
S}={\mathcal S}(\R^d,\R)\to\R$ we require that ${\mathcal C}$ is
continuous wrt the Schwartz topology, normalized, ${\mathcal
C}(0)=1$ and stochastically positive, $\sum_{j,l=1}^n\bar
z_jz_l{\mathcal C}(f_j+f_l)\geq 0$ $\forall$
$n\in\mathbb{N},f_j\in{\mathcal S}, z_j\in\mathbb{C}$. Furthermore,
in order to have a well defined passage from Euclidean time to real
time QFT one requires reflection positivity $\sum_{j,l=1}^n\bar
z_jz_l{\mathcal C}(f_{j,\theta}+f_l)\geq 0$ $\forall$
$n\in\mathbb{N},f_j\in{\mathcal S}_+, z_j\in\mathbb{C}$. Here
${\mathcal S}_+=\{f\in{\mathcal S}: {\rm supp}f\subseteq
\{x\in\R^d:x_1>0\}\}$. Finally, the theory obtained at the boundary
should be conformally invariant, provided the IR cut-off $\Lambda$
is removed from $V_\Lambda$  via taking the limit of the generating
functionals wrt the net $\{\Lambda\}$.

It has been pointed out in \cite{DR,Kn,Re2} that an alternative
representation of the functional (\ref{3.3eqa}) answers a number of
the questions raised above. Let $\phi(\ul{x})=\phi(z,x)$ be the
canonical random field associated with the measure $\mu_{G_+}$. The
idea is to smear $\phi(z,x)$ in the $x$-variable with a test
function $f\in{\mathcal D}(\R^d,\R)$ and then scale $z\to0$. In the
light of (\ref{2.1eqa}), one has to multiply
$\phi(z,f)=\langle\phi,\delta_z\otimes f\rangle$ with a factor
$z^{-\Delta_+}$ in order to obtain a finite result in the limit. We set
 \begin{equation}\label{3.4eqa}
Y_z(f)=\int_{{\mathcal D}'}
e^{\langle\phi,z^{-\Delta_+}\delta_z\otimes
f\rangle}e^{-V_\Lambda(\phi)}\, d\mu_{G_+}(\phi).
\end{equation}
Clearly, under the conditions on $V_\Lambda$ given in the preceding
section and for $\Lambda=\theta\Lambda$, $Y_z(f)/Y_z(0)$ defines a
continuous, normalized, stochastically positive and reflection
positive generating functional for all $z>0$. Using the fact that
$G_+(\delta_z\otimes f)$ is in the Cameron-Martin space of the measure
$\mu_{G_+}$, one gets with $f_z=z^{-\Delta_+}\delta_z\otimes f$, cf.
\cite{GT},
\begin{equation}
\label{3.5eqa} {Y}_z(f)/{Y}_z(0)=
e^{\frac{1}{2}G_+(f_z,f_z)}\int_{\mathcal
D'}e^{-V_\Lambda(\phi+G_+f_z)}d\mu_{G_+}(\phi)/{Y}_z(0).
\end{equation}
We now want to take the limit $z\to 0$. Using (\ref{2.2eqa}) one can
show under rather weak continuity requirements on $V_\Lambda$ that
the functional integral on the rhs of (\ref{3.5eqa}) converges to
$\int_{{\mathcal D}'}e^{-V_\Lambda(\phi+H_+f)} d\mu_{G_+}(\phi)$.
The prefactor however diverges. The reason is that the limit in
(\ref{2.3eqa}) is only a pointwise limit for $x\not=x'$ and not a
limit in the sense of tempered distributions. One can however show
that \cite{GT}
$$\int_{\R^{d}}\int_{\R^{d}}\alpha_+(x,y)f(x)f(y)dxdy=
$$
$$
\lim_{z\rightarrow
0}z^{-2\Delta_+}\int_{\R^{d}}\int_{\R^{d}}G_+(z,x;z,y)f(x)f(y)dxdy -
$$
$$
\frac{1}{(2\pi)^{\frac{d}{2}}}\left(\frac{2^{1-\nu}}{\sqrt{\pi}\Gamma(\nu+\frac{1}{2})}\right)^2\sum_{j=0}^{[\nu]}z^{-2(\nu-j)}(-1)^ja_j\int_{\R^{d}}|\hat{f}(k)|^2|k|^{2j}dk.
$$
\begin{equation}
\label{3.6eqa} =:\lim_{z\rightarrow
0}z^{-2\Delta_+}\int_{\R^{d}}\int_{\R^{d}}G_+(z,x;z,y)f(x)f(y)dxdy
-({\rm Corr}(z)f,f).
\end{equation}
Here $a_j=\int_0^\infty (\int_0^1\cos(\omega
t)(1-t^2)^{\nu-\frac{1}{2}}dt)^2\omega^{2(\nu-j)-1}d\omega$. Thus,
the right hand side of (\ref{3.5eqa}) multiplied with
$e^{-\frac{1}{2}({\rm Corr}(z)f,f)}$ converges and we obtain the
limiting functional
\begin{eqnarray}
\label{3.7eqa} \tilde {\mathcal C}(f)&=&\lim_{z\to
0}e^{-\frac{1}{2}({\rm Corr}(z)f,f)}\left({Y}_z(f)/{Y}_z(0)\right)\nonumber\\
&=& \lim_{z\to 0}e^{\frac{1}{2}[G_+(f_z,f_z)-({\rm
Corr}(z)f,f)]}\int_{\mathcal
D'}e^{-V_\Lambda(\phi+G_+f_z)}d\mu_{G_+}(\phi)/{Y}_z(0)\nonumber\\
&=&e^{\frac{1}{2}\alpha_+(f,f)}\frac{\int_{\mathcal
D'}e^{-V_\Lambda(\phi+H_+f)}d\mu_{G_+}(\phi)}{\int_{\mathcal
D'}e^{-V_\Lambda(\phi)}d\mu_{G_+}(\phi)}
\end{eqnarray}
This, together with $\alpha_{-}^{-1}=-c^2\alpha_+$, establishes the
crucial identity \cite{DR,GT}
\begin{equation}
\label{3.8eqa} {\mathcal C}(f)=\tilde {\mathcal C}(cf),~~\forall
f\in{\mathcal S}(\R^d,\R).
\end{equation}
Let us now investigate the structural properties of the generating
functional ${\mathcal C}: {\mathcal S}\to \mathbb{R}$. If there were
not the correction factor $({\rm Corr}(z)f,f)$, $\mathcal{C}$ would
be stochastically positive and reflection positive as the limit of
functionals with that property, since we can combine (\ref{3.7eqa})
and $(\ref{3.8eqa})$ for a representation of ${\mathcal C}$. However,
due to the signs in (\ref{3.6eqa}) ${\mathcal S}\ni f\to
e^{-\frac{1}{2}({\rm Corr}(z)f,f)}\in\mathbb{\R}$ is not
stochastically positive and consequently the stochastic positivity
of $e^{-\frac{1}{2}({\rm
Corr}(z)f,f)}\left({Y}_z(f)/{Y}_z(0)\right)$ is at least unclear.
Hence we do not have any reason to believe that the limiting
functional ${\mathcal C}$ is stochastically positive and can be
associated with a probabilistic functional integral. An exception is
the case where $V_\Lambda\equiv0$ where we can dwell on the fact
that $\mathcal{S}\ni f\to e^{\frac{1}{2}\alpha_+(f,f)}\in\R$ is
manifestly stochastically positive since $\hat
\alpha(k)=C_{-\nu}\left(\frac{|k|}{2}\right)^{2\nu}\in\mathbb{\R}$
with $C_{-\nu}>0$. It is therefore questionable if one can use the
AdS/CFT correspondence to generate conformally invariant models in
statistical mechanics.

We next investigate the question of reflection positivity. Since the
correlation length of the distributional kernels of ${\rm Corr}(z)$
is zero, we get that $({\rm Corr}(z)(f_{j,\theta}+f_l),(
f_{j,\theta}+f_l))=({\rm Corr}(z)f_{j,\theta},f_{j,\theta})+({\rm
Corr}(z)f_l,f_l)=({\rm Corr}(z) f_j, f_j)+({\rm Corr}(z)f_l,f_l)$
for $f_j\in{\mathcal S}_+$. Consequently, $\forall f_j\in{\mathcal
S}_+,z_1,\ldots,z_n\in \mathbb{C}$ and $\Lambda$ such that
$\theta\Lambda=\Lambda$ we get
\begin{equation}
\label{3.9eqa}
 \sum_{j,l=1}^n{\mathcal C}(f_{j,\theta}+f_l)\bar
z_jz_l= \lim_{z\to 0}\sum_{j,l=1}^n\left(Y_z(c
f_{j,\theta}+cf_l)/Y_z(0)\right)\bar z_j'z_l'\geq 0
\end{equation}
with $z'_j=z_je^{-\frac{1}{2}({\rm Corr}(z)cf_j,cf_j)}$. For a proof
that the reflection positivity of generating functionals implies the
reflection positivity of Schwinger functions \cite{GJ} also in the
absence of stochastic positivity, cf. \cite{Go}.  As in
\cite{DR,GT,Re2}, we thus come to the conclusion that the crucial
property for the existence of a relativistic theory is preserved in
the AdS/CFT correspondence.

Finally we address the invariance properties of the limiting
generating functional ${\mathcal C}$. For being the generating
functional of a CFT, we require invariance under conformal
transformations, i.e. ${\mathcal C}(f)={\mathcal C}(\lambda^{-1}_u
f_u)$ $\forall f\in{\mathcal S}$ where $u$ is an element of the
conformal group on $\R^d$ and
\begin{equation}
\label{3.10eqa} \lambda_u(x)=\left|{\rm det}\left(\frac{\partial
u(x)}{\partial x}\right)\right|^{-\frac{\Delta_+}{d}}.
\end{equation}
Certainly, as long as an interaction with IR cut-off is included in
the definition of ${\mathcal C}={\mathcal C}_\Lambda$, conformal
invariance can not hold. Using the identification of ${\rm
Iso}(\Hy^{d+1})$ and the conformal group on $\R^d$, we get that
$H_+$ intertwines the respective representations on function spaces,
i.e. \cite{GT}
\begin{equation}
\label{3.11eqa} H_+(u(z,x);x')=\left|{\rm det}\left(\frac{\partial
u^{-1}(x')}{\partial
x'}\right)\right|^{\frac{\Delta_+}{d}}H_+(z,x;u^{-1}(x')).
\end{equation}
Combining this, the conformal invariance of $\alpha_+$ under the
given representation of the conformal group and (\ref{3.3eqa}) we
obtain
\begin{equation}
\label{3.12eqa} {\mathcal C}_\Lambda(\lambda_u^{-1}f_u)={\mathcal
C}_{u\Lambda}(f) ~~\forall f\in{\mathcal S}.
\end{equation}
Hence, if the generating functionals $\{{\mathcal C}_\Lambda\}$ have
a unique limit ${\mathcal C}$ wrt the net $\{\Lambda\}$, then
${\mathcal C}$ is reflection positive and conformally invariant and
hence is the generating functional of a boundary CFT.

\section{The Infra-Red Problem and Triviality}
\label{4sect} In this section we investigate the net limit of $\{
{\mathcal C}_\Lambda\}$ which is needed to establish the full
AdS/CFT correspondence. This problem has been left open in \cite{GT}
and we will show that this kind of IR problem behaves somewhat
wired.

The reason is the following: When we identified the generating
functionals ${\mathcal C}_\Lambda$ and $\tilde {\mathcal
C}_\Lambda$, we have seen from the latter functional that it
originated from a usual QFT generating functional with
$z^{-\Delta_+}\delta_z\otimes f$ giving rise to a source term which
needs to be considered in the limit $z\to 0$. As (\ref{3.6eqa})
shows, this source term corresponds to an interaction of an ``exterior
field" with the quantum field $\phi$ which, already for the free
field, has zero expectation but infinite fluctuations in the limit
$z\to 0$. Without any correction term, this would have led to a
generating functional which converges to zero for any $f\not=0$. We
already then needed an ultra-local correction term to deal with the
prescribed infinite energy fluctuations.

If we now switch on the interaction, a shift term $H_+f$ in the bulk
theory is generated, cf. (\ref{3.7eqa}). If we e.g. restrict to
polynomial interactions, this shift leads to re-defined
$f$-dependent couplings that diverge towards the conformal boundary.
This again leads to an an infinite energy transfer and it is
probable that this infinite amount of energy plays havoc with the
generating functional. Here we will show that in some situations
this indeed happens.

Let us first investigate the behavior of the shift $H_+f$ towards
the conformal boundary. Let $f\in{\mathcal S}$ be such that
$f(0)\not=0$. Choosing spherical coordinates, we denote by $f_{\rm
rad}(r)$ the integral of $f(x)$ over the angular coordinates. We get
from (\ref{2.2eqa}) via a change of coordinates
\begin{equation}
\label{4.1eqa}
H_+f(z,0)=\gamma_+ z^{-\Delta_++d}\int_0^\infty\left(\frac{1}{1+r^2}\right)^{\Delta_+}
f_{\rm rad}(zr) r^{d-1} dr
\end{equation}
and we see that the integral on the rhs converges to
$f(0)\int_0^\infty\left(\frac{1}{1+r^2}\right)^{\Delta_+} r^{d-1}
dr=f(0)\times \Gamma(\Delta_+-d/2)\Gamma(d/2)/2\Gamma(\Delta_+) $,
hence $H_+f(z,x)\sim z^{-\Delta_++d}$ if $f(x)\not=0$ by translation
invariance.

Let us now work with the generating functional as defined by
(\ref{3.7eqa}). The prefactor on the rhs is independent of
$\Lambda$, hence we have to investigate the behavior of
\begin{equation}
\label{4.1aeqa} {\mathcal C}'_\Lambda(f)= \frac{\int_{\mathcal
D'}e^{-V_\Lambda(\phi+H_+f)}d\mu_{G_+}(\phi)}{\int_{\mathcal
D'}e^{-V_\Lambda(\phi)}d\mu_{G_+}(\phi)}.
\end{equation}
 We restrict ourselves to
the simplest possible case - an ultra-violet regularized $\phi^4$
potential in arbitrary dimensions $d+1$
\begin{equation}
\label{4.2eqa} V_\Lambda(\phi)=\lambda
\int_\Lambda:\phi_\kappa^4:(\ul{x})\, d_g\ul{x}
\end{equation}
where $\phi_\kappa$ denotes the random field $\phi$ with UV-cut off
$\kappa$. Due to this cut-off, the locality axiom in Section 2 will
in general be violated. This however does not matter in the
following discussion. We furthermore require that
$G_+^\kappa(\ul{x},\ul{x}')=\mathbb{E}[\phi_\kappa(\ul{x})\phi_\kappa(\ul{x}')]$
is a bounded function in $\ul{x}$ and $\ul{x}'$. $\mathbb{E}$ stands
for the expectation wrt $\mu_{G_+}$. The Wick ordering in
(\ref{4.2eqa}) is taken wrt $G_+$, for simplicity. The shifted
potential then is given by
\begin{equation}
\label{4.3eqa} V_\Lambda(\phi+H_+f)=\lambda\int_\Lambda
\sum_{j=0}^4\left({4\atop j}\right):\phi_\kappa^j:(\ul{x})
(H_+f)^{4-j}(\ul{x})\, d_g\ul{x}.
\end{equation}
Taking the expected value of the shifted potential wrt $\mu_{G_+}$,
one obtains $\lambda\times$ $\times \int_\Lambda(H_+f)^4 d_g\ul{x}$
which in the light of (\ref{4.1eqa}) clearly diverges as
$\Lambda\nearrow \Hy^{d+1}$ whenever $f\not=0$.

Let us now focus on a specific class of cut-offs of the form
$\Lambda(z_0)=\Lambda(z_0,l)=[z_0,A]\times[-l,l]^{\times d}$
where we keep $l>0, A >0$ arbitrarily large but fixed. Let
$V(z_0,f)(\phi)=V_{\Lambda(z_0)}(\phi+H_+f)$. Since
$d_g\ul{x}=z^{-d-1}dzdx$ we obtain the scaling of the expected
shifted interaction energy
\begin{eqnarray}
\label{4.4eqa} E(z_0,f)&=&\mathbb{E}[V(z_0,f)]=\lambda
\int_{[z_0,A]}\int_{[-l,l]^{\times d}} (H_+f)^4 (z,x)
\,dx z^{-d-1} dz\nonumber\\
&\sim&   z_0^{-d-4(\Delta_+-d)}~~{\rm as}~~z_0\to0.
\end{eqnarray}
Let us next investigate the fluctuations in the shifted energy as
$z_0\to0$. Denoting the standard deviation of $V(z_0,f)$ with
$\sigma(z_0,f)$, we obtain using (\ref{4.3eqa}) and
$\mathbb{E}[:\phi^a_\kappa:(\ul{x}):\phi^b_\kappa:(\ul{y})]=a!\,\delta_{a,b}G^\kappa_+(\ul{x},\ul{y})^a$,
$a,b\in\mathbb{N}$,
\begin{eqnarray}
\label{4.5eqa}
\sigma(z_0,f)&=&\left[24\int_{\Lambda(z_0)^{\times 2}} G_+^\kappa(\ul{x},\ul{y})^4d_g\ul{x}d_g\ul{y}\right.\nonumber\\
&+&96\int_{\Lambda(z_0)^{\times 2}} H_+f(\ul{x})H_+f(\ul{y})G_+^\kappa(\ul{x},\ul{y})^3d_g\ul{x}d_g\ul{y}\nonumber\\
&+&72\int_{\Lambda(z_0)^{\times 2}} (H_+f)^2(\ul{x})(H_+f)^2(\ul{y})G_+^\kappa(\ul{x},\ul{y})^2d_g\ul{x}d_g\ul{y}\nonumber\\
&+&16\left.\int_{\Lambda(z_0)^{\times 2}}
(H_+f)^3(\ul{x})(H_+f)^3(\ul{y})G_+^\kappa(\ul{x},\ul{y})d_g\ul{x}d_g\ul{y}\right]^{1/2}\\&\sim&
z_0^{-d-3(\Delta_+-d)}~~{\rm or ~slower~as}~~z_0\to 0,\nonumber
\end{eqnarray}
where we took the factors $G_+^\kappa$ out of the integral and
replaced them with a majorizing constant in order to obtain an upper
bound on the scaling. Apparently, the quotient
$\gamma(z_0,f)=2\sigma(z_0,f)/E(z_0,f)\sim z_0^{\Delta_+-d}$ scales
down to zero if $m^2>0$. Using the Chebychev inequality
$\mu_{G_+}(|V(z_0,f)-E(z_0,f)|\leq E(z_0,f)/2)\leq \gamma(z_0,f)^2$
we see from this that $V(z_0,f)\to \infty$ $\mu_{G_+}$-a.s..

To determine the behavior of ${\mathcal C}_{\Lambda(z_0)}'(f)$ for
$f\not=0$ we however need an argument based on the
hypercontractivity estimate $\|F\|_p\leq (p-1)^{n/2}\|F\|_2$
$\forall F$ that are in the $L^p({\mathcal D}',{\mathcal
B},\mu_{G_+})$-closure of the span of Wick monomials
$:\phi(f_1)\cdots\phi(f_s):$ with $s\leq n$. Applying this to
$V(z_0,f)=V_{\Lambda(z_0)}(\phi+H_+f)$ with $n=4$ one obtains
\begin{eqnarray}
\label{4.6eqa}
 \mu_{G_+}\left(V(z_0,f)\leq
\frac{E(z_0,f)}{2}\right)&\leq& \mu_{G_+}\left(|V(z_0,f)-E(z_0,f)|
\geq \frac{E(z_0,f)}{2}\right)\nonumber\\
&\leq&\frac{2^p}{E(z_0,f)^p}\|V(z_0,f)-E(z_0,f)\|_p^p\nonumber\\
&\leq&\frac{2^p}{E(z_0,f)^p}(p-1)^{2p}~\|V(z_0,f)-E(z_0,f)\|_2^{p}\nonumber\\
&=& \gamma(z_0,f) ^p(p-1)^{2p}.
\end{eqnarray}
The next step is to optimize this estimate wrt $p$ for $z_0\to 0$.
Equivalently, one can ask for the minimum of the logarithm
of the rhs wrt to $p$. Taking the
$p$-derivative of this expression and setting it zero yields
$0=\log\gamma(z_0,f)+\frac{2p(z_0)}{p(z_0)-1}+2\log(p(z_0)-1)$ with
$p(z_0)$ the optimal $p$. Apparently, $p(z_0)\to\infty$ as
$z_0\to0$ and thus $2p(z_0)/(p(z_0)-1)\to 2$, hence $p(z_0)$ scales
as
\begin{equation}
\label{4.7eqa} p(z_0)\sim e^{-1}\times\gamma(z_0,f)^{-1/2}\sim C e^{-1} \times
z_0^{-(\Delta_+-d)/2}.
\end{equation}
Combining (\ref{4.6eqa}) and (\ref{4.7eqa}) yields
\begin{eqnarray} \label{4.7aeqa}
& &\mu_{G_+}\left(V(z_0,f)<\frac{E(z_0,f)}{2}\right)\leq  \nonumber\\
& & \gamma(z_0,f)^{e^{-1}\times \gamma(z_0,f)^{-1/2}}\left( e^{-1}\times \gamma(z_0,f)^{-1/2}-1\right)^{2e^{-1}\times \gamma(z_0,f)^{-1/2}}\nonumber \\
& & \sim e^{-2e^{-1}\times \gamma(z_0,f)^{-1/2}} \nonumber \\
& &\sim e^{-2Ce^{-1} \times z_0^{(d-\Delta_+)/2}}
\end{eqnarray}
We have thus seen that the portion of the probability space where
$V(z_0,f)$ does not get large as $z_0\to0$ has a rapidly falling
probability. We need an estimate that controls the negative values
on this exceptional set. The ultra-violet cut-off implies
$:\phi_\kappa^4:(\ul{x})\geq -B c^2_\kappa$, $B$ independent of
$\kappa$, $c_\kappa=\sup_{\ul{x},\ul{y}}|G_\kappa(\ul{x},\ul{y})|$,
$\mu_{G_+}$-a.s., which provides us with a pointwise lower bound for
$V(z_0,f)$ that is depending on $z_0$ as
\begin{equation}
\label{4.8eqa} V(z_0,f)\geq -\lambda B c_\kappa^2|\Lambda(z_0)|=-
[\lambda B c_\kappa^2(2l)^{d}] \times (z_0^{-d}-A^{-d})/d \quad\mu_{G_+}-{\rm
a.s.}
\end{equation}
Combination of (\ref{4.7aeqa}) and (\ref{4.8eqa}) gives for $z_0$
sufficiently small
\begin{equation}
\label{4.9eqa} \mathbb{E}\left[e^{-V(z_0,f)}\right]\leq
e^{-\frac{1}{2}E(z_0,f)}+e^{[\lambda B c_\kappa^2(2l)^{d}]\times (z_0^{-d}-A^{-d})/d
-2Ce^{-1} \times z_0^{(d-\Delta_+)/2} }
\to 0
\end{equation}
 if $\Delta_+> 3d$ $\Leftrightarrow$ $m^2> 6d^2$.
 Furthermore, by Jensen's inequality and $\mathbb{E}[V(z_0,0)]=0$,
 \begin{equation}
 \label{4.10eqa}
\mathbb{E}[e^{-V(z_0,0)}]\geq e^{-\mathbb{E}[V(z_0,0)]}=1,
\end{equation}
\label{4.11eqa} which implies that for $m^2$ sufficiently large
\begin{equation}
\label{4.12eqa}
\mathcal{C}'_{\Lambda(z_0)}(f)=\frac{\mathbb{E}[e^{-V(z_0,f)}]}{\mathbb{E}[e^{-V(z_0,0)}]}\to
0~~{\rm as}~~z_0\to 0,
\end{equation}
We have thus obtained the following result:

\begin{thm} If the the generating functional ${\mathcal
C}(f)=\lim_{\Lambda}\mathcal{C}_\Lambda(f)$ exists for the
UV-regularized $:\phi^4:$-interaction and is unique (as required in
order to obtain conformal invariance from AdS-invariance) it is also
trivial ($\mathcal{C}(f)=0$ if $f\not=0$) provided $m^2\geq
6d^2$.
\end{thm}
\noindent
The above triviality result relies on three crucial assumptions.
\begin{itemize}
\item[(i)] The potential is quartic, cf. (\ref{4.2eqa});
\item[(ii)] There is a UV-cut-off;
\item[(iii)] The mass is sufficiently large.
\end{itemize}
In order to assess the relevance of the triviality result for the
general case, let us give some short comments on the role of each of
these assumptions:

(i) At the cost of a more restrictive mass bound, assumption (i) can
easily be relaxed from quartic to polynomial interactions. For
non-polynomial interactions, however, the hypercontractivity
estimate can not be used. This might be of relevance, if we consider
$V$ as an effective potential, which in general will be non
polynomial.

(ii) The fact that there is a UV-cut-off enters our triviality
argument via (\ref{4.8eqa}). When removing the UV-cut-off at least
in dimension $d+1=2$, we therefore have to modify the triviality
argument. It turns out that the bound obtained from the
hypercontractivity estimate \cite{GJ,GT} for the UV-problem is not
good enough to reproduce the above argument. It seems to be
necessary to combine UV and IR - hypercontractivity bounds in a
single estimate in order to obtain triviality without cut-offs in
$d+1$ dimensions. We will come back to this point elsewhere.

(iii) The mass bound to us rather seems to be a technical
consequence of the methods used and not so much a true necessity for
the onset of triviality. Different methods, e.g. based on
decoupling via Dirichlet- and Neumann boundary conditions on a
partition of $\Hy^{d+1}$ \cite{GJ} e.g. combined with large
deviation methods might very well lead to less restrictive  mass
bounds or eliminate them completely.

On a heuristic level, the problem that expectation and variance of
the shifted potential and the non shifted potential will have
different scalings under the limit $\Lambda\nearrow\Hy^{d+1}$
prevails for a large class of polynomial and non-polynomial
interactions with and without cut-offs. Thus, in our eyes, the three
assumptions (i)--(iii) are not essential but rather technical. The
result above therefore should be taken rather as an example of what
can happen in the AdS/CFT correspondence than a definite
mathematical statement. Of course, at the present and very
preliminary state of the affair, everybody is free to think
differently.

\section{Conclusions and Outlook}
In this section we give an essentially non-technical discussion on
repair strategies that would cure the obstacle of triviality.

(i) coupling constant renormalization: The simplest way to deal with
the divergences in the potential energy $V(z_0,f)$ would be to make
$\lambda$ a $z_0$-dependent quantity. In fact, a naive guess at the
scaling behavior suggests that $\lambda(z_0)\sim
z_0^{d+4(\Delta_+-d)}$ would compensate for the increase in the
expected value of the interaction energy $V(z_0,f)$ such that with
the modified coupling $\lim_{z_0\to0}\mathbb{E}[V(z_0,f)]=\lambda
C\int_{\R^d}f^4\, dx$ converges to a constant with
$C=\left(\gamma_+\Gamma(\Delta_+-d/2)\Gamma(d/2)/2\Gamma(\Delta_+)\right)^4$,
cf. (\ref{4.1eqa}) and the paragraph thereafter. Furthermore, one
can expect that the subleading terms ($j=1\ldots4$ in
(\ref{4.3eqa})) converge to zero and do not affect the generating
functional. It thus seems reasonable that with this renormalization
the generating functional gives in the limit $z_0\to 0$
\begin{equation}
\label{5.1eqa} \mathcal{C}(f)=e^{\frac{1}{2}\alpha_+(f,f)-\lambda
C\int_{\R^d}f^4\, dx}
\end{equation}
which is reflection positive as a limit of reflection positive
functionals (it is manifestly not stochastically positive for all
$\lambda>0$ and hence gives a nice illustration for the destruction
of stochastic positivity due to the correction term in
(\ref{3.6eqa}) and (\ref{3.7eqa}). The problem with this functional
however is that the additional term in the interaction is an ultra
local term and hence does not influence the corresponding real time
CFT -- which is a free theory determined by the analytic
continuation of $\alpha_+$. Hence this sort of renormalization only
trades in another kind of triviality for the triviality observed in
Section 4.

(ii) bulk counterterms: Such terms can simply be added to the
(formal) Lagrangian. The problem to use this method in the AdS/CFT
correspondence is twofold: Firstly, the infra-red divergences that
are occurring in $V(z_0,f)$ are $f$-dependent. If we however want to
cure them with $f$-dependent counterterms, the renormalization
description of ${\mathcal C}_{\Lambda(z_0)}$ becomes $f$-dependent.
Bulk counterterms however only preserve the structural properties of
stochastic and reflection positivity, if the same renormalization
prescription is chosen for all $f$. Hence, $f$-dependent counterterms would lead to a limiting functional, for which it is not
known, whether it is reflection positive or not. The situation is
worsened from the observation that, unlike in other IR problems, in
the AdS/CFT correspondence the divergences in the nominator and
denominator scale differently - as seen in our triviality result.
This means for bulk counterterms, that, if they are working out fine
for the nominator, they probably create new divergences in the
denominator. Different renormalizations for the potential in the
nominator and in the denominator in the limit might lead to a non
normalizable vacuum for the boundary theory, which does not make
sense.

(iii) boundary counterterms: The problems described above for bulk
counterterms also have to be taken into account for boundary
counterterms. Furthermore, while bulk counterterms, at least if they
are not $f$-dependent, do not spoil the conformal invariance of the
boundary theory, boundary counterterms theoretically might do so.
Hence one needs a separate argument to show that they don't. But
there is still another problem with boundary counterterms. We have
seen that we can not take it for granted that a limiting functional
measure exists for the boundary theory. But if the boundary theory
is not described by a functional integral $\mu_{\rm bd.}$, it is not
clear how to define boundary counterterms on a mathematical basis:
recall that a counterterm (at a finite value of the cut-off $z_0$ is
defined by $d\mu_{\rm bd.,ren,z_0}(\varphi)=e^{-{\mathcal L}_{\rm
ren}(z_0,\varphi)}d\mu_{\rm bd.}(\varphi)/\int_{\mathcal{D}(\R^d)}
e^{-{\mathcal L}_{\rm ren}(z_0,\varphi')}d\mu_{\rm bd.}(\varphi')$
and it is not obvious how this can be defined if $\mu_{\rm bd.}$ is
not a measure.

(iv) giving up generating functionals: The triviality result of
Section 4 relied on the scaling behavior of the expected value of
$V(z_0,f)$ under the limit $z_0\to 0$. This expected value can be
associated with the Witten graph $\bigotimes$ which gives rise to
the first order contribution to the four point function
$\int_{\Hy^{d+1}}\prod_{l=1}^4H_+(\ul{x},f_l)\, d_g\ul{x}$ which is
converging as long as ${\rm supp} f_j\cap{\rm supp} f_l=\emptyset$
if $j\not=l$, cf (\ref{2.2eqa}) and (\ref{4.1eqa}) (see also
\cite{Kn} for concrete calculations). One may thus hope that the
triviallity result of Section 4 is an artefact of using generating
functionals which makes it necessary to evaluate Schwinger functions
at unphysical coinciding points. A reasonable approach to the infra-red problem in AdS/CFT would thus be to use (\ref{3.7eqa}) to define
reflection positive Schwinger functions with cut off and then remove
the cut-off for the Schwinger functions at physical (non coinciding)
points, only. This might then work out without further
renormalization along the lines of \cite{GJ}, as divergences might
only occur on the diagonal. If this is true, triviality does only
occur on the level of generating functionals -- which are
reminiscent of the Laplace transform of a functional measure for the
boundary theory that might not exist in the present context.

\label{7sect}


\subsection*{Acknowledgment}
The authors gratefully acknowledge interesting discussions with
Sergio Albeverio, Matthias Blau, Michael D\"utsch and Gordon Ritter.
The first named author also would like to thank the organizers of
the conference ``Recent developments in QFT" for creating a very
interesting meeting.

\begin{thebibliography}{1}

\bibitem{AJPS} Albeverio, S., Jost, J., Paycha, S., Scarlatti, S, A Mathematical Introduction to String Theory: Variational Problems, Geometric and Probabilistic
Methods, London Mathematical Society Lecture Note Series, Cambridge
University Press 1997.

\bibitem{BBMS} Bertola M., Bros J., Moschella U., Schaeffer R.: Decomposing quantum fields on branes. Nuclear Physics {\bf B 581}, 575-603 (2000)

\bibitem{BEM} Bros J., Epstein H., Moschella U.: Towards a general theory of quantized fields on the anti-de Sitter spacetime. Commun. Math. Phys. {\bf 231}, 481-528 (2002)


\bibitem{DR}
D\"utsch M., Rehren K.H.: A comment on the dual field in the AdS-CFT
correspondence. Lett. Math. Phys. {\bf 62}, 171-184 (2002)



\bibitem{GJ}
Glimm J., Jaffe A.: {\it Quantum Physics. A Functional Integral
Point of View}, 2nd edition. Springer, New-York 1987

\bibitem{GT} Gottschalk, H., Thaler, H.: {\it AdS/CFT correspondence in the Euclidean
context}, math-ph/0611006, to appear in Commun. Math. Phys.

\bibitem{Go}
Gottschalk H.: {\it Die Momente gefalteten Gau\ss -Poissonschen
wei\ss en Rauschens als Schwingerfunktionen}. Diploma thesis, Bochum
1995

\bibitem{GKP}

Gubser S.S., Klebanov I.R., Polyakov A.M.: Gauge theory correlators
from noncritical string theory. Phys. Lett. {\bf B 428}, 105-114
(1998)

\bibitem{Ha}
Haba Z.: Quantum field theory on manifolds with a boundary. J. Phys.
{\bf A 38}, 10393-10401 (2005)


\bibitem{JR} Jaffee A., Ritter G.: Quantum field theory on curved
backgrounds II: Spacetime symmetries. arXiv:0704.0052v1 [hep-th]

\bibitem{Kn}
Kniemeyer O.: {\it Untersuchungen am erzeugenden Funktional der
AdS-CFT-Korre}\-{\it spondenz}. Diploma thesis, Univ. G\"ottingen 2002


\bibitem{Ma}
Maldacena J.: The large N limit of superconformal field theories and
supergravity. Adv. Theor. Math. Phys. {\bf 2}, 231-252 (1998)

\bibitem{OS1}
Osterwalder K., Schrader R.: Axioms for Euclidean Green's functions.
Comm. Math. Phys. {\bf 31}, 83-112 (1973)

\bibitem{OS2}
Osterwalder K., Schrader R.: Axioms for Euclidean Green's functions.
II. With an appendix by Stephen Summers.  Comm. Math. Phys. {\bf
42},  281--305 (1975)

\bibitem{Re1}
Rehren K.-H.: Algebraic holography. Ann. Henri Poincar\`{e} {\bf 1},
607-623 (2000)

\bibitem{Re2}
Rehren, K.-H.: QFT lectures on AdS-CFT, arXiv:hep-th/0411086v1


\bibitem{Wi}
Witten E.: Anti-de Sitter space and holography. Adv. Theor. Math.
Phys. {\bf 2}, 253-291 (1998)

\end{thebibliography}
\end{document}